\documentclass[twocolumn]{aastex631}

\usepackage{threeparttable}  

\begin{document}


\title{Red Spiral Galaxies at Cosmic Noon Unveiled in the First JWST Image}

\author[0000-0001-7440-8832]{Yoshinobu Fudamoto} 
\affiliation{Waseda Research Institute for Science and Engineering, Faculty of Science and Engineering, Waseda University, 3-4-1 Okubo, Shinjuku, Tokyo 169-8555, Japan}
\affiliation{National Astronomical Observatory of Japan, 2-21-1, Osawa, Mitaka, Tokyo, Japan}

\author[0000-0002-7779-8677]{Akio K. Inoue} 
\affiliation{Waseda Research Institute for Science and Engineering, Faculty of Science and Engineering, Waseda University, 3-4-1 Okubo, Shinjuku, Tokyo 169-8555, Japan}
\affiliation{Department of Physics, School of Advanced Science and Engineering, Faculty of Science and Engineering, Waseda University, 3-4-1, Okubo, Shinjuku, Tokyo 169-8555, Japan}

\author[0000-0001-6958-7856]{Yuma Sugahara} 
\affiliation{Waseda Research Institute for Science and Engineering, Faculty of Science and Engineering, Waseda University, 3-4-1 Okubo, Shinjuku, Tokyo 169-8555, Japan}
\affiliation{National Astronomical Observatory of Japan, 2-21-1, Osawa, Mitaka, Tokyo, Japan}




\begin{abstract}
In the first image of the James Webb Space Telescope (JWST) of SMACS J0723.3-7327,
one of the most outstanding features is the emergence of a large number of red spiral galaxies,
because such red spiral galaxies are only a few percent in the number fraction among nearby spiral galaxies.
While these apparently red galaxies were already detected with the Spitzer Space Telescope at $\sim3-4{\rm \mu m}$, the revolutionized view from JWST's unprecedented spatial resolution has unveiled their hidden spiral morphology for the first time.
Within the red spiral galaxies, we focus on the two reddest galaxies that are very faint in the $<0.9\,{\rm \mu m}$ bands and show red colors in the $2-4\,{\rm \mu m}$ bands.
Our study finds that the two extremely red spiral galaxies are likely to be in the Cosmic Noon ($1 < z < 3$). 
One of the extremely red spiral galaxies is more likely to be a passive galaxy having moderate dust reddening (i.e., $\sim$zero star formation rate with $\rm{A_{V}\sim1\,mag}$).
The other is consistent with both passive and dusty starburst solutions (i.e., star formation rate $> 100\,\rm{M_{\odot}\,yr^{-2}}$ with $\rm{A_{V}\sim3\,mag}$)
These ``red spiral'' galaxies would be interesting, potentially new population of galaxies, as we start to see their detailed morphology using JWST, for the first time.
\end{abstract}



\section{Introduction} \label{sec:intro}
The spiral structure of galaxies is not only one of the most spectacular features of the Universe but also provides us with important information of galaxy formation and evolution.
Since the first systematic classification of the morphology of ``extragalactic nebulae'' \citep{Hubble26}, large efforts have been devoted to studying the morphology of galaxies across cosmic time and to understanding their formation mechanisms \citep[see][for a review]{Conselice14}.
However, when and how the galaxy morphology emerged in the early Universe is still largely unknown.

Spiral galaxies typically show blue colors in their rest-frame optical wavelength and are, in general, classified as ``normal'' star-forming galaxies (e.g., \citealt{2014MNRAS.440..889S}).
Red or passive (i.e. non star-forming or `anemic'; \citealt{1976ApJ...206..883V}) spiral galaxies are, on the other hand, a very minor population in the nearby Universe.
In the latest study, \cite{2022PASJ...74..612S} identified nearly a thousand red, passive spiral galaxies among $\sim55,000$ galaxies at $0.01<z<0.3$ from $1000\,{\rm deg^2}$ imaging data obtained with the Subaru/Hyper Suprime-Cam. Hence, the fraction of the red spiral galaxy is only $\sim2\%$ in the local Universe.

It was, for the first look, surprising to find many apparently red spiral galaxies in the James Webb Space Telescope (JWST) image of the galaxy cluster, SMACS J0723.3-7327, that was released on July 11th 2022 as part of the early release observations \citep[ERO;][]{Pontoppidan22}.
Their redness could indicate several important properties of these spiral galaxies: their dominant stellar ages, dust reddenings, or a combination of these features.
Because the released image was taken by the Near Infrared Camera (NIRCam) on JWST over the wavelength range of $0.9-4.4\,{\rm \mu m}$, the redness in NIR may indicate that these spiral galaxies are at high redshift.

Spiral galaxies in the distant universe are a very important population to examine the emergence of the spiral structure in galaxies.
The most distant spiral galaxy known so far is a gas-rich galaxy, BRI 1335-0417, at $z=4.41$ \citep{2021Sci...372.1201T}.
A grand-design spiral structure traced by the [C~{\sc ii}] line was revealed with the Atacama Large Millimeter/submillimeter Array (ALMA); however, the stellar structure of BRI 1335-0417 is still unknown.
The most distant, spiral stellar disk galaxies are reported at $z=2$--3 \citep{2003AJ....125.1236D,2012Natur.487..338L,2022MNRAS.511.1502M,Wu2022}. 
Searching for such galaxies at even higher redshift is the key to knowing when the stellar spiral disks emerged.

To characterize red spiral galaxies found in the JWST's ERO data, we carefully select the most extreme red spiral galaxies and examine their spectral energy distribution (SED), as a first case study.
This Letter is organized as follows: in \S2 we describe data and the sample used in this study. In \S3, we present our analysis.  \S4 shows the results and discussion on the red spiral galaxies found in the JWST images. We conclude in \S5.
Throughout this Letter, we assume a cosmology with $(\Omega_m,\Omega_{\Lambda},h)=(0.3,0.7,0.7)$

\section{Data} \label{sec:data}
The following images of the Hubble Space Telescope (HST) and JWST are based on a ``first-pass'' reduction of the HST and JWST images of the SMACS-0723 lensing cluster field. All images have been processed with the grizli\footnote{\url{https://github.com/gbrammer/grizli}} software pipeline.  Further documentation will be provided by G. Brammer et al. (in prep).

\subsection{HST data}
As part of the Reionization Lensing Cluster Survey: RELICS \citep{Coe19}, optical and NIR images were obtained using HST.
These HST images include data from ACS (F435W, F606W, F814W filters) and WFC3/IR (F105W, F125W, F140W, F160W filters) instruments. All HST data were calibrated and mosaiced in a standard manner. Final images have a pixel scale of $0.04^{\prime\prime}/{\rm pixel}$.

\subsection{JWST data}
We used JWST NIRCam images of six filters in total (F090W, F150W, F200W, F277W, F356W, and F444W).
After the re-processing of the JWST's public data available in the MAST data archive, the final mosaic has $0.02^{\prime\prime}/{\rm pixel}$ for the short wavelength filters (F090W, F150W, F200W), and $0.04^{\prime\prime}/{\rm pixel}$ for the long wavelength filters (F277W, F356W, F444W).
A small astrometric offset ($\sim 0.2^{\prime\prime}$) exists between HST and JWST images.
The offset, however, did not affect our analysis as we manually corrected positions of apertures during our photometry.

\begin{figure}[tb]
    \centering
    \includegraphics[width=0.7\columnwidth]{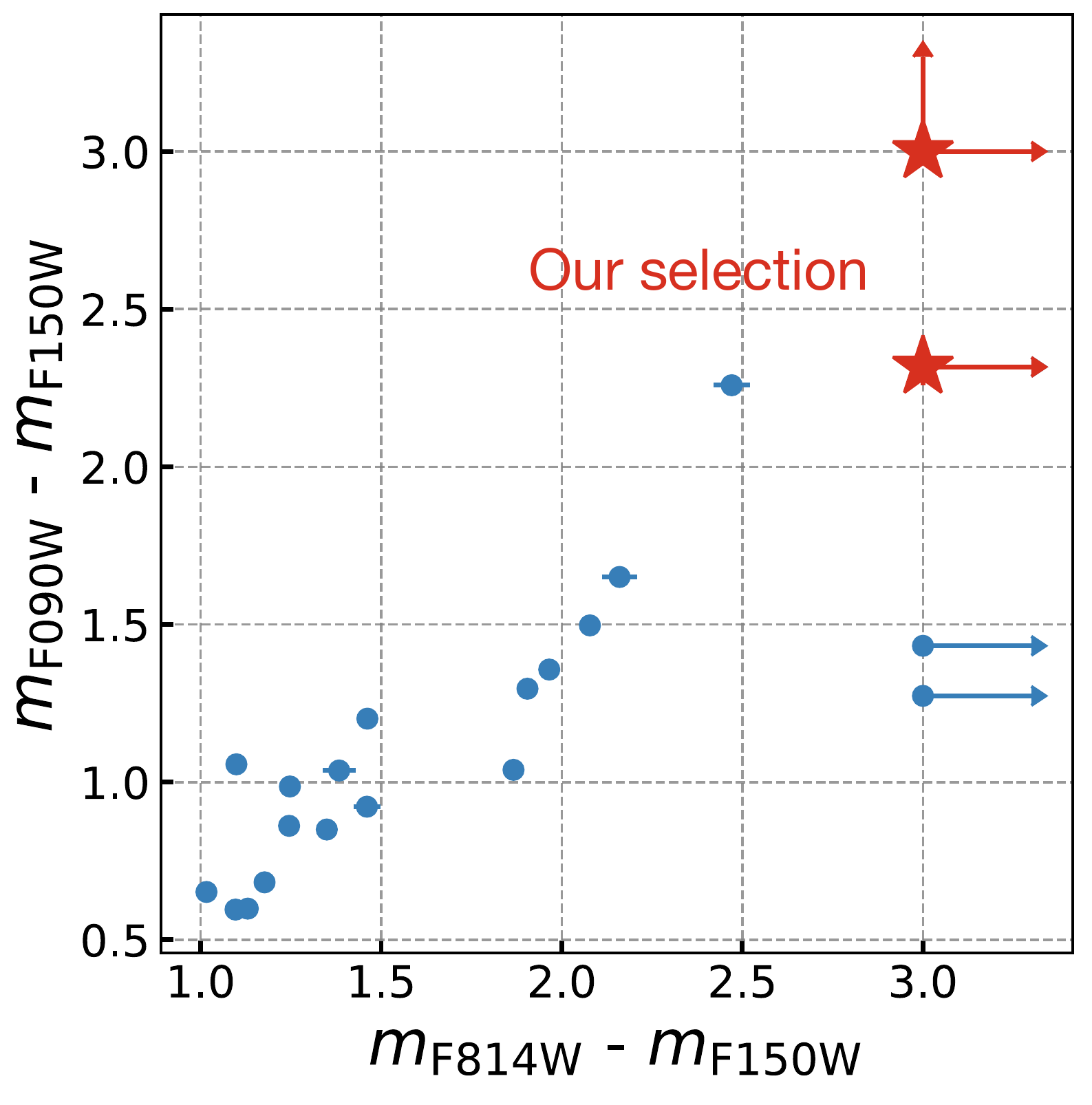}
    \caption{The observed colors of red spiral galaxies. After our visual selections of red spiral galaxies, we further down-selected the most $\sim1-2\,{\rm \mu m}$ red galaxies, which we focus on in this study (red stars). These galaxies have extremely red colors of $m_{\rm F814W/F090W}-m_{\rm F150W} > 2.5\,{\rm mag}$. Other red spiral galaxies (blue points) have at least one $>3\,\sigma$ detection in the HST ACS bands. We applied zero-point corrections for JWST's fluxes based on \citet{Adams2022}}.
    \label{fig:colors}
\end{figure}

\begin{figure*}[tb]
    \centering
	\includegraphics[width=0.6\textwidth]{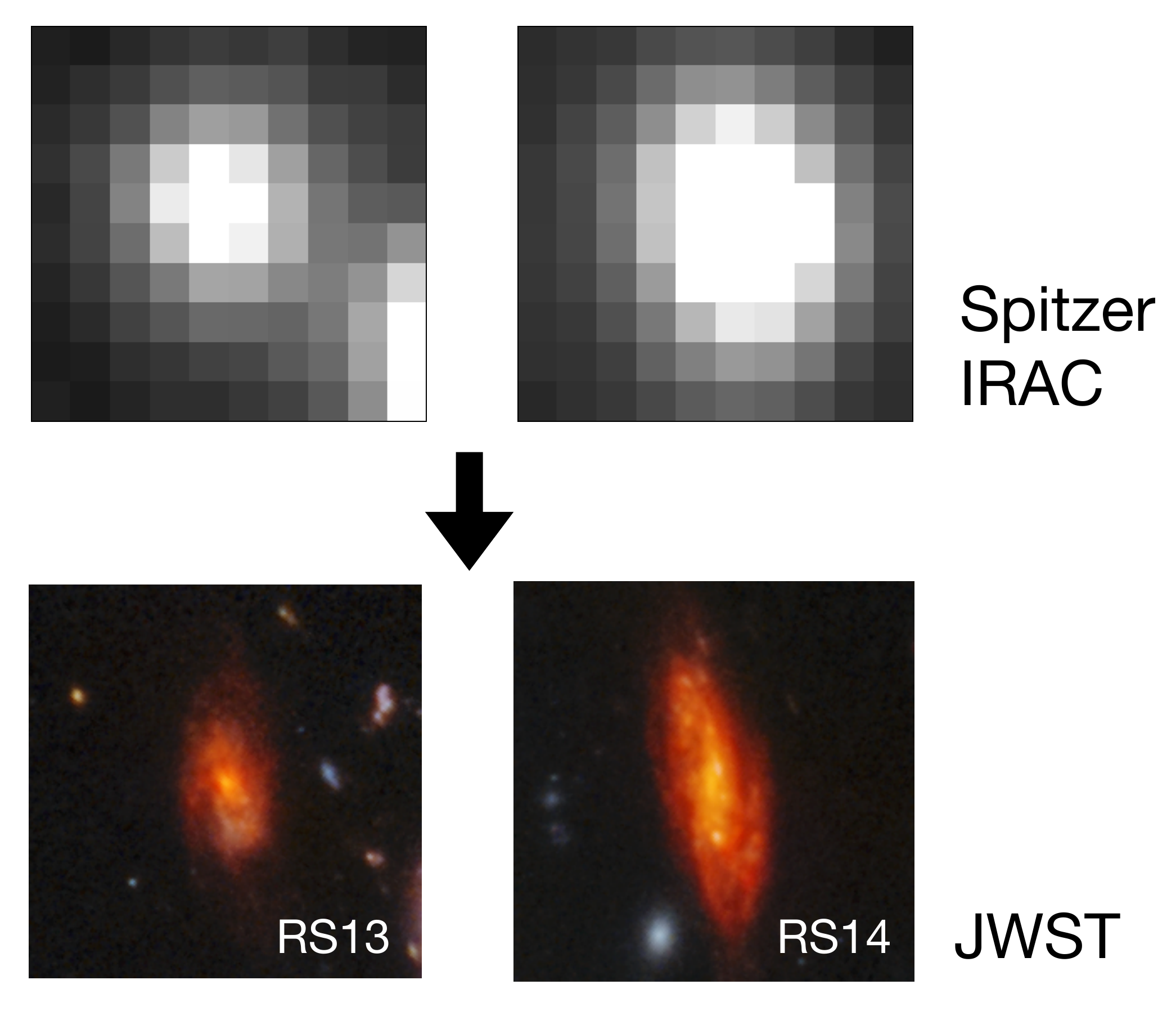}
    \caption{ Demonstrations of the improved spatial resolution by JWST observations: (Upper panels) Spitzer IRAC $3.6\,{\rm \mu m}$ images of the extremely red spiral galaxies in the SMACS 0723 field.
    (Lower Panels) False-color images by JWST's F090W, F150W, F200W, F277W, F356W, and F444W filters.
    All images have spatial scales of $5^{\prime\prime}\times5^{\prime\prime}$.
    The arm and bulge structure of the galaxies are captured for the first time, thanks to the extremely high resolution and sensitivity of JWST. North is up and East is left for both Spitzer IRAC and JWST images.  Copyright: NASA/STScI}
    \label{fig:nasaimage}
\end{figure*}

\section{Analysis} \label{sec:analysis}

\subsection{Selecting Red Spiral Galaxies}
We first visually selected bright red spiral galaxies from the false-color image presented within the ERO release\footnote{\url{https://stsci-opo.org/STScI-01G7DDBW5NNXTJV8PGHB0465QP.png}}. 
Two authors (Y.F. and A.K.I.) visually selected red spiral galaxies from the image independently.
We selected 21 galaxies that both agreed as having spiral structures with apparently red colors.
While quantitative evaluation of galaxy types would also be preferred \citep[e.g.,][]{Dominguez18,Ferreira22}, visual selections are also known to be effective and widely used, especially when finding apparent structures, such as spiral shape \citep[e.g., Galaxy Zoo Project:][]{Lintott08}.

To find the reddest spiral galaxies, we further down-selected samples based on their faintness in $\leq0.9\,{\rm \mu m}$ images (see \S\ref{sec:photometry} below for the photometry of our sample here). 
In particular, we selected galaxies that are non-detected in the HST F814W image and, at the same time, galaxies that have an extremely red color in the F090W vs. F150W image (Fig. \ref{fig:colors}).

The selected red spiral galaxies (RS13 and RS14) were already detected by HST WFC3 and Spitzer IRAC images; however, the angular resolution of these previous instruments did not allow us to study detailed morphology. 
We can now access the resolved morphology of the red spiral galaxies by JWST's unprecedented spatial resolution and sensitivity in these NIR wavelengths (Fig. \ref{fig:nasaimage}).

\subsection{Photometry}

\subsubsection{HST and JWST photometry}
\label{sec:photometry}
To measure galaxy-wide integrated fluxes, we performed aperture photometry of both HST and JWST images for the red spiral galaxies.
We used python package \texttt{photoutils} \citep{larry_bradley_2020_4044744} using elliptical apertures.
We applied apertures large enough to enclose entire galaxies in the JWST F444W image to obtain integrated fluxes.
Measured flux errors incorporated background noises by scaling pixel-by-pixel root mean squares (RMSs) to aperture sizes, and Poisson noises of measured fluxes. 

\subsubsection{JWST photometric uncertainty}
As calibrations of NIRCam data progressed, studies reported that the NIRCam's photometric zero points have uncertainties, which could be up to $\sim20\,\%$ different from the pre-flight measurements \citep{Adams2022,Morishita22,Rigby22} and/or could be time-variable \citep{Nardiello2022}. In this study, we use the photometric zero-point of JWST NIRCam data based on \citet{Adams2022}, which is derived using updated JWST's calibration references for the SMACS 0723 field used using in-flight data.
These updated zero-points were, however, based on preliminary calibrations of JWST. Therefore, further uncertainties could still be expected. To incorporate such uncertainties, we additionally applied $10\%$ of flux uncertainties for all measured fluxes from JWST. These uncertainties were added quadratically to the measured aperture flux errors. Additionally, we tested the effect of changing photometric zero points in the following discussions, including values derived from in-flight throughput in \citep{Rigby22} and the original pre-flight values.

\subsubsection{RS14}
Within the extremely red spiral galaxies, RS14 has already been studied in several papers.
In particular, \citet{Sun22}  reported ALMA continuum detection in $1.1\,{\rm mm}$ image obtained among the ALMA large program ALCS: 2018.1.00035.L (PI: K. Kohno). Most recently, \citet{Cheng22} presented a joint JWST-ALMA study of this source.
Also, \citet{Carnall22} reported spectroscopic redshift of $z_{\rm spec}=2.463$ determined by multiple emission lines such as H$\alpha$, [NII], and [SII] emission lines (ID 9239 of NIRSpec spectroscopy as part of the same JWST ERO).
In the following analysis, we used the ALMA continuum flux and the NIRSpec redshift for RS14.

\subsection{Lensing Magnification}
We estimated gravitational lensing magnification based on the recent work by \citet{Golubchik22}.
Their lens model is constructed with \texttt{Light-Traces-Mass} method \citep[e.g.,][]{Broadhurst05}, which infers the lens mass contributions from the cluster's light distributions, using HST and VLT MUSE data. We estimated lensing magnification factors by assuming redshifts from the following SED fittings.

\subsection{Spectral Energy Distribution fittings}
\label{sec:SEDfit_results}

To constrain the physical properties and redshift of RS13, one of our extremely red spiral galaxies, we used SED fitting code \texttt{PANHIT} \citep{Mawatari20}. 
We assumed a Chabrier initial mass function in a range of 0.1--100 M$_\odot$ \citep{2003PASP..115..763C}, the BC03 stellar population model \citep{2003MNRAS.344.1000B}, a nebular continuum and line emission model \citep{2011MNRAS.415.2920I}, and a standard dust attenuation curve \citep{Calzetti00}.
The star formation history was assumed to be the ``delayed-$\tau$'' model \citep{2014ApJS..214...15S} to describe a variety of star formation histories, and the star-formation time-scale $\tau_{\rm SF}$ is a free parameter between 0.01 and 10 Gyr.
We allowed for metallicities ranging from $0.005\times{Z_{\odot}}$ to $2.5\times{Z_{\odot}}$, where the Solar metallicity of $Z_\odot=0.02$. 
The dust attenuation in the $V$-band ($A_{V}$) is also a free parameter. 
The age of the stellar population is another free parameter between 1 Myr and 15 Gyr, and the cosmic age limits it at the redshift of interest.
For RS13, the redshift is another free parameter and we examined redshifts from 0.1 to 13.0 with a step of 0.1.
The stellar mass and star formation rate (SFR) were determined by scaling the amplitude of the SED.
The SED fits performed $\chi^{2}$ minimization algorithm for the photometry data, including non-detection bands \citep[i.e., upper limits;][]{Sawicki12}.
Additionally, to test robustness of our SED fittings, we also used SMC dust extinction curve\citep{Gordon2003} as well as assuming constant and then truncate star formation history.

For RS14, we run the fitting with the spectroscopic redshift of $z=2.463$ \citep[][]{Carnall22} and including the ALMA continuum flux \citep{Cheng22}. 
{\tt PANHIT} assumes energy conservation between the stellar radiation absorbed by dust and the far-infrared (FIR) emission from dust.
We assumed modified black-body functions with an emissivity index $\beta=2$ and the dust temperature, $T_{\rm d}$, between 10 and 50 K with a step of 5 K.
Note that the cosmic microwave background temperature at $z=2.463$ is 9.5 K, a strict lower limit of $T_{\rm d}$.

\begin{figure*}[bt]
    \centering
    \includegraphics[width=0.8\textwidth]{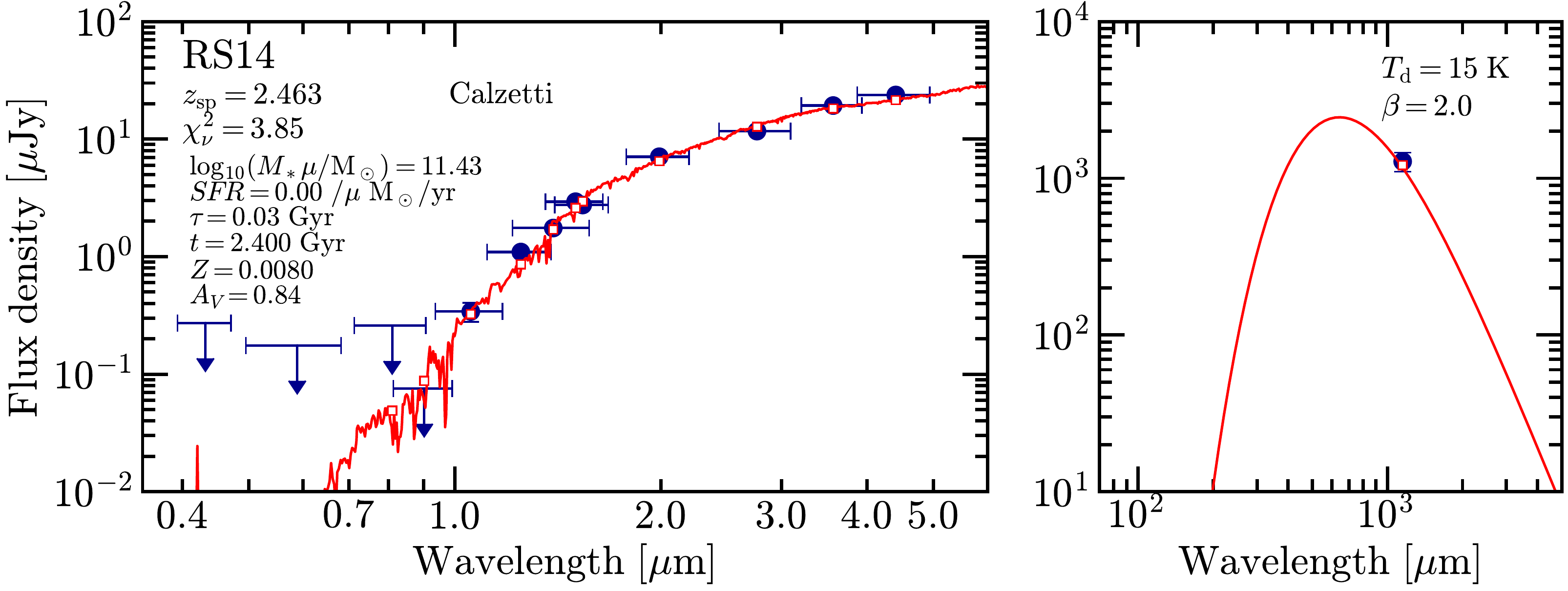}
    \caption{The best fit SED of RS14 derived by fixing its redshift at $z_{\rm sp}=2.463$. The integrated JWST's photometry can be represented solely by old stellar populations, thus not requiring considerable dust reddening and star formation. The SED fitting results indicate that RS14 represents a passive spiral galaxy at high redshift.}
    \label{fig:SED_RS14}
\end{figure*}

\begin{figure*}
    \centering
    \includegraphics[width=0.8\textwidth]{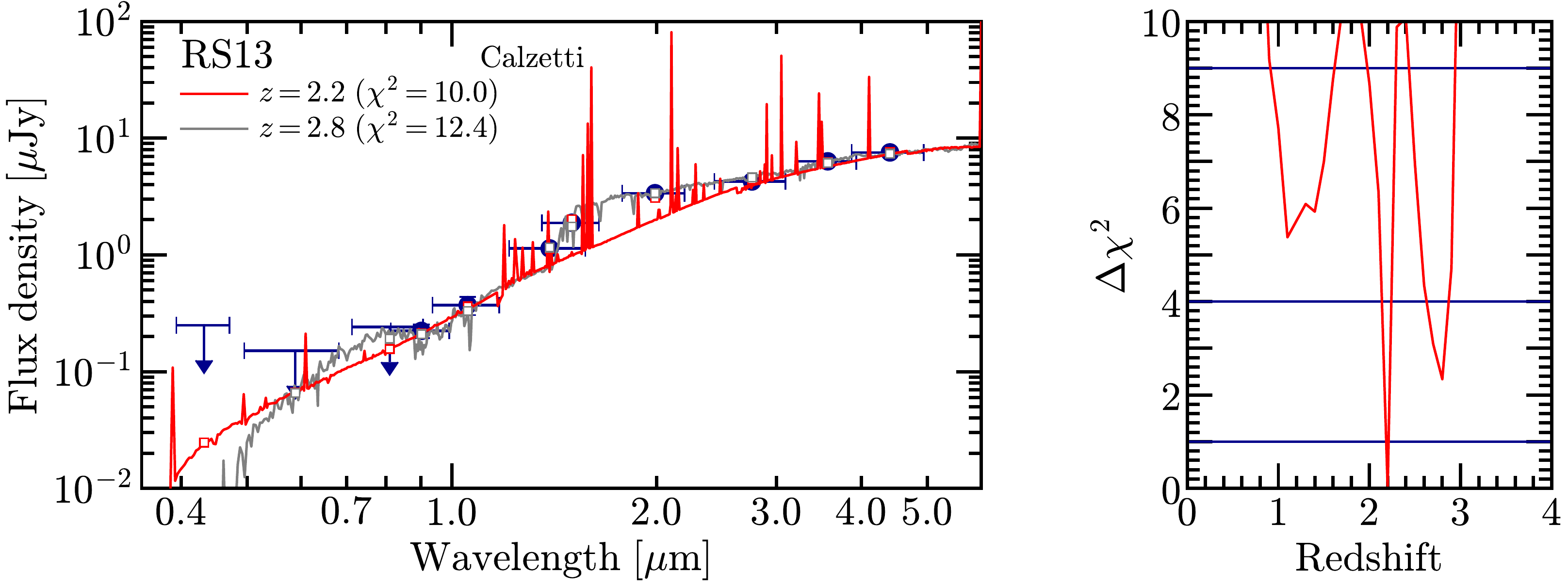}
    \caption{SED fitting results of RS13: Left panel shows the SEDs inferred from our fitting. Right panel shows reduced $\chi^{2}_\nu$ as a function of redshift. 
    For RS13, photometric redshift is loosely constrained at $1<z<3$.
    The best-fit is a dusty star forming solution at $z_{\rm ph}=2.2\pm0.1$ (red solid line in the left panel), and second best-fit is a passive solution with SFR$\sim0$ at $z_{\rm ph}=2.8^{+0.1}_{-0.2}$ (gray solid line in the left panel).
    }
    \label{fig:SED_RS1213}
\end{figure*}

\section{Results and Discussion} \label{sec:results}

\subsection{SED fitting reuslts}

We found that the redshifts of extremely red spirals are in a range of $z_{\rm ph}=1$--3.
Especially, RS14 ($z_{\rm spec}=2.463$) is one of the highest redshift spiral galaxies identified so far \citep[][]{2003AJ....125.1236D,2012Natur.487..338L,2022MNRAS.511.1502M,Wu2022}.

Our SED fits showed that two types of integrated properties exist for the extremely red spiral galaxies at $1<z<3$: (1) Old stellar population that is consistent with almost completely passive, non star-forming galaxies, or (2) Dust-obscured highly elevated star-formation activities. In our SED fits, these properties are indicated from the JWST's coverage of $\lambda_{\rm rest}\sim4300 - 12800\,\text{\AA}$ photometry.
The red colors of these bands can be explained by old stellar populations with moderate dust attenuation, or  dust obscured star-formation activity.
Especially, the possibility of a population of red and dead spiral galaxies at $z>1$ in our extremely red spiral sample is very interesting, as they are very rare in the local Universe \citep{2022PASJ...74..612S}.

We note that our SED fit results could have some caveats, such as uncertain calibration of JWST photometry, which would change observed colors of these galaxies \citep[see also][for different photometric zero-point corrections]{Morishita22}. Also, the physical properties of these red spiral galaxies are not well studied yet. Thus, other assumptions, that are not included in our SED fits, could be allowed; such as extremely steeper dust attenuation curve.
However, we test several reasonable cases to incorporate these possible caveats. As a result,  we always find that  the extremely red spiral galaxies show photometric redshift $z\sim1-3$ and have passive stellar population or dust obscured star formation activities. In particular, RS14 always show passive stellar population even if we use several photometric zero points or assumptions for SED fits.

Table~\ref{tab:sedresults} is a summary of the fitting results as well as the lensing magnification factors. 
Following, we describe results of our SED fitting in detail:

\renewcommand{\arraystretch}{0.85}
\begin{table}[h]
\centering
\caption{A summary of SED fitting results and lensing magnification factors.}
\label{tab:sedresults}
    \begin{tabular}{ccccc}
    \hline
    RS13 & best fit\\
    \hline
    $z_{\rm ph}$ & $2.2\pm0.1$ & $2.8^{+0.1}_{-0.2}$ \\
    $\chi^2_\nu$ & 1.25 & 1.54 \\
    $\log_{10}(M_* \times \mu ~[{\rm M_\odot}])$ & $9.95^{+0.07}_{-0.01}$ & $10.45_{-0.01}^{+0.04}$\\
    $SFR \times \mu$ (M$_\odot$ yr$^{-1}$) & $450\pm170$ & $0.17_{-0.17}^{+0.01}$\\
    $A_V$ (mag) & $3.07_{-0.07}^{+0.06}$ & $0.59_{-0.20}^{+0.13}$\\
    Age (Gyr) & $0.020_{-0.002}^{+0.025}$ & $0.36_{-0.09}^{+0.13}$ \\
    $\tau_{\rm SF}$ (Gyr) & 0.01--10 & $0.03_{-0.02}^{+0.01}$ \\
    $Z$ & 0.03--0.06 & 0.03--0.05 \\
    $\mu$ & $1.67\pm0.02$ & $1.75\pm0.03$ \\
    \hline
    RS14\\
    \hline
    $z_{\rm sp}$ & 2.463 & \\
    $\chi^2_\nu$ & 3.84 \\
    $\log_{10}(M_* \times \mu ~[{\rm M_\odot}])$ & $11.43\pm0.03$\\
    $SFR \times \mu$ (M$_\odot$ yr$^{-1}$) & 0.00 \\
    $A_V$ (mag) & $0.84_{-0.03}^{+0.12}$ \\
    Age (Gyr) & $2.4_{-0.2}^{+0.1}$ \\
    $\tau_{\rm SF}$ (Gyr) & 0.03--0.1 \\
    $Z$ & 0.007--0.012 \\
    $T_{\rm dust}$ (K) & $\sim15$ \\
    $\mu$ & $2.16\pm0.06$ & \\
    \hline
\end{tabular}
\end{table}

\subsubsection{RS14}
We found that the $T_{\rm d}=15$ K case delivered the minimum $\chi^2$ value compared to other $T_{\rm d}$ cases.
Higher $T_{\rm d}$ with fixing 1.1 mm flux leads to larger IR luminosity, and thus, larger dust attenuation; however, it makes the rest-frame optical SED traced by NIRCam too red. As such, the $T_{\rm d}=15$ K, rather low temperature \citep[e.g.,][]{2018A&A...609A..30S}, seems the best representation for RS14, that is shown in Fig.~\ref{fig:SED_RS14}.

The best-fit model for RS14 is passive, old stellar population, and moderate dust attenuation ($A_{\rm V}=0.84^{+0.12}_{-0.03}$).
The age is much longer than the star-formation time-scale, $\tau_{\rm SF}\lesssim0.1\,{\rm Gyr}$, and the current SFR is almost zero. This feature support the very low $T_{\rm d}$ ($15\,{\rm K}$) that is heated by old stars rather than massive young stars.
The stellar mass is relatively high, $1.0\times10^{11}\,{\rm M_{\odot}}$ after correcting a magnification of $\mu=2.16$ at $z_{\rm spec}=2.463$. 

The best-fit passive solution of the RS14 does not change if we use different photometric zero point, including pre-flight value, or if we use different assumptions for the SED fitting. In particular, we also applied a steeper dust attenuation (i.e., the one from SMC dust extinction curve; \citealt{Gordon2003}). However the results do not change.

While the galaxy-wide integrated property of RS14 is consistent with non-star-forming passive galaxy, JWST's F090W image (the rest-frame wavelength of $\sim2600\,\text{\AA}$) shows a few faint clumps in the bulge and some parts of the spiral arm region of RS14, which is smoothed out by our photometry using a large aperture. As these clumps can be seen in the rest-frame UV wavelength, a small amount of star-formation activity is still on-going in RS14.
This is also consistent with the H$\alpha$ emission line detection \citep{Carnall22}.
Hence, RS14 may have a small but substantial star-formation activity.

\subsubsection{RS13}

For RS13, we found a global $\chi^{2}$ minimum in the redshift range of $1\lesssim z\lesssim3$ (lower right panel of Fig. \ref{fig:SED_RS1213}), showing two local minima at $z=2.2$ and $z=2.8$. 

The best redshift solution of RS13 is $z_{\rm ph}=2.2\pm0.1$, having a large dust reddening of $A_{\rm V}=3.07^{+0.06}_{-0.07}$ with a high SFR of $\sim450\,\rm{M_{\odot}\,yr^{-1}}$. This solution explains some of the $>1\,\rm{\mu m}$ fluxes seen in NIRCam bands are produced by strong optical-to-NIR nebular lines, instead of the stellar continuum. With the young age of $0.02\,\rm{Gyr}$, this solution suggests that RS13 is a dusty starburst spiral galaxy.

Another redshift solution is $z_{\rm ph}=2.8^{+0.1}_{-0.2}$. 
For this solution, RS13 is, similar to RS14, a passive galaxy (SFR consistent with zero) with the stellar age of $0.4\,{\rm Gyr}$, showing the clear Balmer break around F140W and F150W filter wavelengths. The stellar mass is $\sim2\times10^{10}\,{\rm M_{\odot}}$ after correcting a magnification.

Although these two solutions show largely different properties for  RS13, both solution shows that the redshift of RS13 to be $1<z<3$. This general results do not change if we use different photometric zero points or different assumptions for SED fittings.
Further disentangling these two possible properties (i.e., a dusty starburst spiral or a passive spiral) requires spectroscopic redshift.

\section{Conclusion}
In this paper, we studied the properties of two extremely red spiral galaxies (RS13 and RS14) found from the JWST ERO image data release of SMACS 0723.
These extremely red spiral galaxies are among a sample of red spiral galaxies visually selected from the ERO data.
By performing SED fitting to these extremely red spiral galaxies, we found  following results:

\noindent$\bullet$ Most likely redshifts of the extremely red spirals are $1 < z < 3$, i.e., in the Cosmic Noon. The results show that these red spiral galaxies in JWST images contain the most distant spiral galaxies known to date. Within them, RS14 is currently one of the most distant stellar-spiral galaxy, having $z_{\rm spec}=2.463$.

\noindent$\bullet$ Our SED fits suggest that the extremely red spiral galaxies are passive galaxies or heavily dust-obscured galaxies.
These properties are indicated from the red colors of the rest-frame wavelength of $\sim4300 - 12800\,\text{\AA}$ in the JWST photometry.

\noindent$\bullet$ One of the extremely red spiral galaxies, RS14, is found to be passive spiral galaxy.
Finding  passive spiral galaxies in the early Universe is surprising, as most of the spiral galaxies found in the local Universe is young star forming galaxies.

Further detailed studies would be required to understand the formation mechanisms and evolutionary path of the red spiral galaxies.  Follow-up observations using integral field spectroscopy, including high-resolution ALMA observations, will provide kinematics, molecular gas, and dust distribution of high-redshift spiral galaxies.

\begin{acknowledgments}
We thank Fengwu Sun, Ryosuke Uematsu, and Marc Postman for very helpful discussions.
Y.F., A.K.I., and Y.S. acknowledge support from NAOJ ALMA Scientific Research Grant number 2020-16B. 
The Early Release Observations and associated materials were developed, executed, and compiled by the ERO production team:  Hannah Braun, Claire Blome, Matthew Brown, Margaret Carruthers, Dan Coe, Joseph DePasquale, Nestor Espinoza, Macarena Garcia Marin, Karl Gordon, Alaina Henry, Leah Hustak, Andi James, Ann Jenkins, Anton Koekemoer, Stephanie LaMassa, David Law, Alexandra Lockwood, Amaya Moro-Martin, Susan Mullally, Alyssa Pagan, Dani Player, Klaus Pontoppidan, Charles Proffitt, Christine Pulliam, Leah Ramsay, Swara Ravindranath, Neill Reid, Massimo Robberto, Elena Sabbi, Leonardo Ubeda. The EROs were also made possible by the foundational efforts and support from the JWST instruments, STScI planning and scheduling, and Data Management teams.
\end{acknowledgments}

%

\vspace{5mm}
\facilities{HST, JWST (STIS)}


\software{astropy \citep{astropy:2013,astropy:2018},  
          photoutils \citep{larry_bradley_2020_4044744},
          PANHIT \citep{Mawatari20}
          }


\bibliography{RedSpirals}{}
\bibliographystyle{aasjournal}

\end{document}